\documentclass[prd,preprint,nofootinbib]{revtex4}
\usepackage{graphicx}
\usepackage{amssymb}
\usepackage{amsmath}

\begin{document}

\title{CMB constraints on the simultaneous variation of the fine structure constant and electron mass}
\author{Kazuhide  Ichikawa}
\author{Toru Kanzaki}
\author{Masahiro Kawasaki}
\affiliation{Institute for Cosmic Ray Research,
University of Tokyo,
Kashiwa 277-8582, Japan
}
\date{\today}

\begin{abstract}
We study constraints on time variation of 
the fine structure constant $\alpha$ from cosmic microwave background (CMB) 
taking into account simultaneous change in
$\alpha$ and the electron mass $m_e$ which might be implied in unification theories. 
We obtain the constraints $-0.097 < \Delta \alpha/\alpha < 0.034$ 
at 95$\%$ C.L. using WMAP data only, and
$-0.042 < \Delta \alpha/\alpha < 0.026$
combining with the constraint on the Hubble parameter by the HST Hubble Key Project.
These are improved by 15\% compared with constraints assuming only $\alpha$ varies. We discuss other relations between variations in $\alpha$ and $m_e$ but we do not find evidence for varying $\alpha$.
\end{abstract}

\maketitle

\section{INTRODUCTION }\label{sec:intro}

One of fundamental questions in physics is whether or not 
the physical constants are literally constant. 
In fact, the physical constants may change in spacetime within the context of some 
unification theories such as superstring theory and investigating their constancy is an important probe of those theories. 
In addition to the theoretical possibility, some observations suggest the time variation of the fine structure constant (the coupling constant of electromagnetic interaction) $\alpha$. Recent constraints including such non-null results are briefly summarized as follows.

Terrestrial limits on $\alpha$ come from atomic 
clocks~\cite{Prestage1995,Bize2003,Marion2003,Fischer2004,Peik2004,Bize2005,Peik2005}, the Oklo natural fission reactor in 
Gabon~\cite{Fujii:2002hc} and meteorites. Ref.~\cite{Peik2005} derived the limit on the temporal derivative of $\alpha$ at present as $(-0.3 \pm 2.0) \times 10^{-15}$yr$^{-1}$. Measurements of Sm isotopes in the Oklo provide two bounds on the variation of $\alpha$ as 
$\Delta \alpha/\alpha = -(0.8 \pm 1.0) \times 10^{-8}$ and 
$\Delta \alpha/\alpha = (0.88 \pm 0.07) \times 10^{-7}$ \cite{Fujii:2002hc}. 
The former result is null, however, the latter is a strong detection 
that $\alpha$ was larger at $z \sim 0.1$. The meteorite bound obtained by measuring $^{187}$Re decay rate is now controversial although varying $\alpha$ is not suggested anyway. Ref.~\cite{Olive:2003sq} obtained $\Delta\alpha/\alpha = (-8\pm8)\times 10^{-7}$, whereas Refs.~\cite{Fujii:2003uw,Fujii:2005wq} argued that the constraint should be much weaker due to uncertainties in the decay rate modeling. 

On the other hand, there are three kinds of celestial probes. One
is to use big bang nucleosynthesis (BBN) \cite{Bergstrom:1999wm,Ichikawa:2002bt,Nollett:2002da,Ichikawa:2004ju} which provides constraints at very high redshifts ($z
\sim 10^9$-$10^{10}$), for example, $-5.0 \times 10^{-2} < \Delta \alpha/\alpha < 1.0 \times 10^{-2}$ (95\% C.L.) \cite{Ichikawa:2002bt} or $|\Delta \alpha/\alpha| < 6 \times 10^{-2}$ \cite{Cyburt:2004yc}.
 The second is from the spectra of high-redshift quasars ($z \sim 1$-$3$) \cite{Webb:2000mn,Murphy:2003hw,Srianand:2004mq,Levshakov:2004bg,Chand:2004et,Kanekar:2005xy,Levshakov:2005ab}, where there are conflicting results.
  Refs.~\cite{Webb:2000mn,Murphy:2003hw} suggested that $\alpha$ was smaller at $z \gtrsim 1$, $\Delta \alpha/\alpha= (-0.543 \pm 0.116) \times 10^{-5}$ \cite{Murphy:2003hw}.
   This result, however, is not supported by other observations \cite{Srianand:2004mq,Levshakov:2004bg,Chand:2004et,Kanekar:2005xy,Levshakov:2005ab}. For example, Ref.~\cite{Srianand:2004mq} obtained the constraint as
$\Delta \alpha/\alpha = (-0.6 \pm 0.6) \times 10^{-6}$ and the others too found no evidence for varying $\alpha$. Finally, we can use Cosmic Microwave Background (CMB) to measure $\alpha$ at $z\sim 1100$ \cite{Hannestad:1998xp,Kaplinghat:1998ry,Battye:2000ds,Landau:2000dd}. Analyses using pre-WMAP data are found in Refs.~\cite{Battye:2000ds,Avelino:2000ea,Landau:2000dd}. Refs.~\cite{Martins:2003pe,Rocha:2003gc} derive a constraint using the WMAP first-year data, $-0.05 < \Delta\alpha/\alpha < 0.02$ or $-0.06 < \Delta \alpha/\alpha < 0.01$ (95\% C.L.) respectively with or without marginalizing over the running of the spectral index \cite{Rocha:2003gc}.
\footnote{
Note that two results suggesting varying $\alpha$, one of the Oklo results \cite{Fujii:2002hc} and the quasar observation of Ref.~\cite{Murphy:2003hw} have different signs of $\Delta \alpha$. These results, if correct, can not be explained by homogeneous and monotonically time-varying $\alpha$. They may indicate that $\alpha$ is not a monotonically varying function of time or as investigated in Refs.~\cite{Barrow:2002zh,Mota:2003tc,Mota:2003tm}, may suggest a spatial variation of $\alpha$. In passing, we refer to Ref.~\cite{Sigurdson:2003pd} for the effects of spatial varying $\alpha$ on CMB. 
}

Recently, there are many studies on constraining the time variation in $\alpha$ which accompanies the variation of the other coupling constants, as would occur rather naturally in unified theories. Under such a framework, BBN has been studied in Refs.~\cite{Campbell:1994bf,Langacker:2001td,Dent:2001ga,Calmet:2001nu,Ichikawa:2002bt,Calmet:2002ja,Olive:2002tz,Muller:2004gu,Landau:2004rj}, quasar absorption systems in Ref.~\cite{Flambaum:2004tm}, meteorites in Ref.~\cite{Olive:2003sq}, the Oklo reactor in Ref.~\cite{Olive:2002tz}, and atomic clock experiments in Refs.~\cite{Flambaum:2004tm,Flambaum:2006ip}. For example, BBN constraint improves by up to about 2 orders ($-6.0 \times 10^{-4} < \Delta\alpha/\alpha < 1.5 \times 10^{-4}$ \cite{Ichikawa:2002bt}) although the factor may vary depending on how they are correlated to $\alpha$. Hence, it is important to consider the variation of other coupling constants along with $\alpha$.  

In this paper, we investigate constraint on the time variation of $\alpha$ from CMB using the WMAP first-year data. In particular, we consider $m_e$ to vary dependently to the variation in $\alpha$ because, as mentioned above, such might be the case in some unified theory. Since such theory is now under development, how their variations are related to each other can not be predicted. Therefore, we work in a phenomenological way guided by a low energy effective theory of a string theory and adopt to vary $m_e$ in power law of $\alpha$.
\footnote{
Refs.~\cite{Kujat:1999rk,Yoo:2002vw} studied CMB constraints on the variation of Higgs expectation value whose effect is assumed to appear only in the change in $m_e$ so there is some overlap between their analysis and ours. However, they did not consider $\alpha$ variation at the same time.
}

In the next section, we briefly review the recombination process in the early universe and make clear how it depends on $\alpha$ and $m_e$. In section \ref{sec:CMB}, we illustrate their effects on the epoch of recombination and the shape of CMB power spectrum. In section \ref{sec:relation}, we describe the relations between the variations in $\alpha$ and $m_e$ which we adopt in our analysis. In section \ref{sec:constraint}, we present our constraint and we conclude in section \ref{sec:conclusion}.

\section{$\alpha$ and $m_e$ dependence of the recombination process}     \label{sec:rec}
Non-standard values of $\alpha$ and $m_e$ modify the CMB angular power 
spectrum  mainly by changing the epoch of recombination.  
Thus, let us visit briefly the recombination process in the universe and see where those constants appear. We follow the treatment of Ref.~\cite{Seager:1999bc}, which is implemented in the RECFAST code. They have shown that the recombination process is well approximated by the evolutions of three variables: the proton fraction $x_p$, the singly ionized helium fraction $x_{\rm HeII}$, and the matter temperature $T_M$. Their equations are given bellow. We denote the Boltzmann constant $k$, the Planck constant $h$ and the speed of light $c$. In addition to the variables above, $x_p = n_p/n_{\rm H}$ and $x_{\rm HeII} =n _{\rm HeII}/n_{\rm H}$, we use the electron fraction $x_e = n_e/n_{\rm H} = x_p +x_{\rm HeII}$ as an auxiliary variable (note that $n_X$ stands for the number density of species $X$ but $n_{\rm H}$ is defined as the total hydrogen number density, including both protons and hydrogen atoms). $z$ is used for the redshift and $H$ for the expansion rate.

Adopting the three level approximation, the time evolution of 
the proton fraction of $x_p$ is described by
\begin{eqnarray}
& &\hspace{-0.5cm}\frac{dx_p}{dz} = \nonumber \\
 & &  \frac{C_{\rm H}}{H(z)(1 + z)} 
    \left[ x_e x_p n_{\rm H} R_{\rm H} -\beta_{\rm H} (1 - x_p) e^{-h\nu_{\rm H}/kT_M}\right],
    \label{eq:proton_ratio} \nonumber \\
\end{eqnarray}
where $\nu_{\rm H} = c/(121.5682\ {\rm nm})$ is the Ly$\alpha$ frequency and $R_{\rm H}$ is the case B recombination coefficient for H which is well fitted by
\begin{eqnarray}
 R_{\rm H} &= & 10^{-19}F\frac{at^b}{1 + ct^d}~\mathrm{m^3s^{-1}}
       \label{eq:recom_H}  
\end{eqnarray}
 with $t= T_M/(10^4\ {\rm K})$, $a= 4.309$, $b = -0.6166$, $c = 0.6703$, $d = 0.5300$ \cite{Pequignot}, and the fudge factor $F = 1.14$ introduced to reproduce the more precise multi-level calculation \cite{Seager:1999bc}. $\beta_{\rm H}$ is the photoionization  coefficient
\begin{eqnarray}
\beta_{\rm H} = R_{\rm_H} \left(\frac{2\pi m_e k T_M}{
h^2}\right)^{\frac{3}{2}} \exp \left(-\frac{B_{\rm H2s}}{kT_M}\right),
\end{eqnarray}
and $C_H$ is the so-called Peebles reduction factor
\begin{eqnarray}
 C_{\rm H} =  \frac{\left[1 + K_{\rm H}\Lambda_{\rm H} n_{\rm H}(1 - x_p) \right]}{\left[1 + K_{\rm H}(\Lambda_{\rm H} + \beta_{\rm H})
         n_{\rm H}(1 - x_p) \right]}, \label{eq:reduction_factor}
     \label{eq:peebles-supp} 
\end{eqnarray}
where the binding energy in the 2s energy level is $B_{\rm H2s} = 3.4$ eV, the two-photon decay rate is $\Lambda_{\rm H} = 8.22458$ ${\rm s}^{-1}$ and $K_{\rm H} = c^3/(8\pi\nu_{\rm H}^3 H)$.

Here, comments on what Eq.~(\ref{eq:proton_ratio}) means may be in order. The first term in the square brackets in Eq.~(\ref{eq:proton_ratio})
represents the recombinations to excited states of the atom, ignoring
recombination direct to the ground state.  The second term represents
the rate of ionization from excited states of the atom.  The
difference of those two terms is the net rate of production of hydrogen
atoms when one could ignore the Ly$\alpha$ resonance photons.  These
photons reduce the rate by the factor $C_{\rm H}$, which can be written as the ratio of the net decay rate to the sum of the decay
and ionization rates from the $n = 2$ level,
\begin{equation}\label{eq:reduction_factor_rewrited}
        C_{\rm H} = \frac{\Lambda_{\rm R} + \Lambda_{\rm H}}
        {\Lambda_{\rm R} + \Lambda_{\rm H} + \beta_{\rm H}}.
\end{equation}
We have rewritten Eq.~(\ref{eq:reduction_factor}) to derive this expression using the decay rate
$\Lambda_{\rm R} \equiv (K_{\rm H}n_{1s})^{-1}$ allowed by redshifting
of Ly$\alpha$ photons out of the line, and using $n_{1s} = n_{\rm H}-n_p$ which is justified by the far greater occupation number of the hydrogen atom ground state than that of the excited states altogether.

The evolution of the singly ionized helium fraction of $x_{\rm HeII}$ is similarly described by
\begin{eqnarray}
\frac{dx_{\rm HeII}}{dz} &=& \frac{C_{\rm He}}{H(z)(1 + z)}  \nonumber \\
& &\hspace{-1.5cm}\times\left[x_{\rm HeII}x_e n_{\rm H} R_{\rm HeI}- \beta_{\rm HeI}(f_{\rm He} - x_{\rm HeII}) e^{-h\nu_{\rm HeI}/kT_M}]\right], \nonumber \\
\end{eqnarray}     
where $f_{\rm He}$ is the total number fraction of
helium to hydrogen (using primordial helium mass fraction $Y_p$, $f_{\rm He} = Y_p/\{4(1-Y_p)\}$ where we take $Y_p$ to be 0.24), $\nu_{\rm HeI} = c/(60.1404\ {\rm nm})$ is the frequency corresponding to energy 
between ground state and $2^1 s$ state, and $R_{\rm HeI}$ is the case B HeI recombination 
coefficient for singlets \cite{Hummer}
\begin{eqnarray}
        R_{\rm HeI} &=& \nonumber \\
& &\hspace{-1cm}q\left[\sqrt{\frac{T_M}{T_2}} \left(1 + \frac{T_M}{T_2}\right)^{1 -p} \left(1 + \frac{T_M}{T_1}\right)^{1 + p}\right]^{-1} ~\mathrm{m^3s^{-1}}, \nonumber \\
        \label{eq:recom_He}
\end{eqnarray}
with $q = 10^{-16.744}$, $p = 0.711$,
$T_1 = 10^{5.114}{\rm K}$ and $T_2$ fixed arbitrary 3K. $\beta_{\rm HeI}$ is the photoionization  coefficient
\begin{eqnarray}
\beta_{\rm HeI} = R_{\rm HeI} \left(\frac{2\pi m_e k T_M}{
h^2}\right)^{\frac{3}{2}} \exp \left(-\frac{B_{\rm HeI2s}}{kT_M}\right),
\end{eqnarray}
\begin{eqnarray}
     & &   C_{\rm HeII} = \nonumber \\
 & &  \hspace{-0.5cm}  \frac{\left[ 1 + K_{\rm HeI}\Lambda_{\rm He}
        n_{\rm H}(f_{\rm He} - x_{\rm HeII})\exp(\Delta E/kT_M)\right]}
        {\left [1 + K_{\rm HeI}(\Lambda_{\rm He} + \beta_{\rm HeI})n_{\rm H}
        (f_{\rm He} - x_{\rm HeII})\exp(\Delta E/kT_M)\right]}, \nonumber \\
\end{eqnarray}
where the binding energy in the 2s energy level is $B_{\rm HeI2s} = 3.97$ eV and 
the two-photon decay rate is $\Lambda_{\rm He} = 51.3\ {\rm s}^{-1}$, $K_{\rm He I} = c^3/(8\pi\nu_{\rm He I}^3H)$ and $\Delta E$ is the energy separation between $2^1 s$ and $2^1 p$, $\Delta E/h = c/(58.4334\ {\rm nm}) - \nu_{\rm HeI} $.
Note that, contrary to the case with H, the energy separation $\Delta E$ between $2^1 s$ and $2^1 p$ is so large that we can not neglect it \cite{Matsuda}.

The matter temperature $T_M$ is evolved as
\begin{eqnarray}
& & \hspace{-0.5cm} \frac{dT_M}{dz} = \nonumber \\
& &\frac{8\sigma_Ta_R{T_R}^4}{3H(z)(1 + z)m_e}
     \frac{x_e}{1 + f_{\rm He} + x_e}(T_M - T_R) + \frac{2T_M}{(1 + z)}, \nonumber \\
\end{eqnarray}
where $T_R$ is the radiation temperature, $\sigma_T=2\alpha^2 h^2/(3\pi m_e^2c^2)$ 
is the Thomson cross section, and $a_R = k^4/(120\pi c^3 h^3)$ is
the black-body constant. 
The Compton scattering makes $T_R$ and $T_M$ 
identical at high redshifts. However, the adiabatic cooling 
becomes dominant at low redshifts, which leads to the significant 
difference between $T_M$ and $T_R$. 

Now, we explain how quantities which appear in these equations depend on $\alpha$ and $m_e$. Two-photon decay rates scale as $\alpha^8m_e$~\cite{Breit,Uzan:2002vq}. Since binding energies scale as $\alpha^2 m_e$, so do $\nu_{\rm H}$ and $\nu_{\rm HeI}$, and $K_{\rm H}$ and $K_{\rm HeI} $ scale as $\alpha^{-6}m_e^{-3}$.
The remaining task is to investigate how the recombination
coefficient $R$ depends on 
$\alpha$ and $m_e$. To do this, we follow the treatment of 
Ref.~\cite{Kaplinghat:1998ry}. 
The recombination coefficient can be expressed as 
\begin{eqnarray}
        R  & = & \sum_{n,l}^{*}[8\pi(2l + 1)]
        \left(\frac{kT_M}{2\pi m_e}\right)^{\frac{3}{2}}
        \exp\left(\frac{B_n}{kT_M}\right)
        \nonumber \\
        & & \times \int_{B_n/kT_M}^{\infty} 
        \frac{\sigma_{nl} y^2dy}{\exp(y) - 1},
\end{eqnarray}
where $B_n$ is the binding energy for n-th excited state 
and $\sigma_{nl}$ is the ionization cross section for ($n$,$l$) 
excited state. The asterisk in the upper bound of summation indicates that the sum needs 
to be regulated, but since this regularization depends only weakly 
on $\alpha$ and $m_e$, it can be neglected~\cite{Boschan:1996hu}. 
The cross section $\sigma_{nl}$ scales as 
$\alpha^{-1}m_e^{-2}$~\cite{Uzan:2002vq}. Altogether,
\begin{eqnarray}
        \frac{\partial R(T_M)}{\partial \alpha} &= & 
        \frac{2}{\alpha}\left(R(T_M) 
        - T_M\frac{\partial R(T_M)}{\partial T_M}\right),        \\
        \frac{\partial R(T_M)}{\partial m_e} &= & 
        - \frac{1}{m_e}\left(2R(T_M) 
        + T_M\frac{\partial R(T_M)}{\partial T_M}\right).
\end{eqnarray}
Combining with the fitting formulae (\ref{eq:recom_H}) and (\ref{eq:recom_He}), 
we obtain how $R_{\rm H}$ and $R_{\rm HeI}$ depend on $\alpha$ and $m_e$.

\begin{figure}
  \centering
  \includegraphics[width=7.5cm]{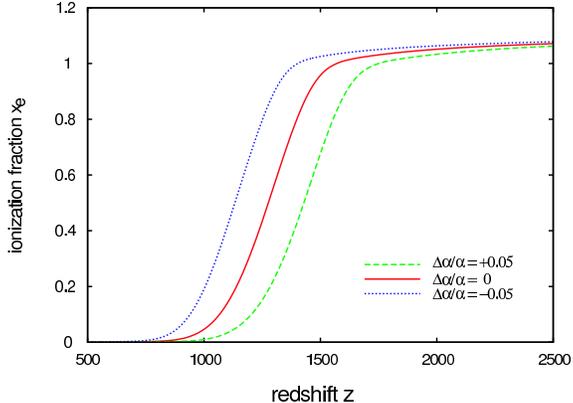}
  \caption{The ionization fraction $x_e$ as a function of 
  redshift $z$ for no change of $\alpha$ (solid curve), 
  an increase of $\alpha$ by 5$\%$ (dashed curve), a decrease of 
  $\alpha$ by 5$\%$ (dotted curve).}
  \label{fig:a_ion_frac}
\end{figure}

\begin{figure}
   \centering
   \includegraphics[width=7.5cm]{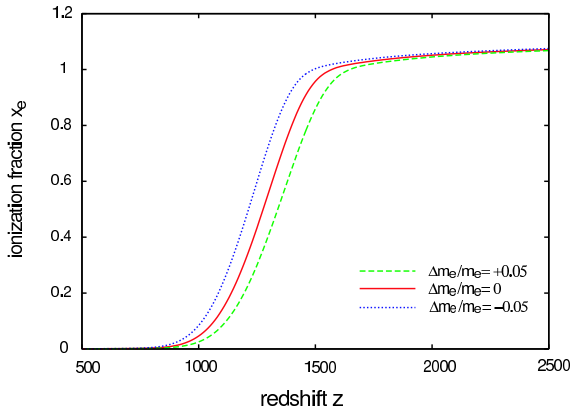}
   \caption{The ionization fraction $x_e$ as a function of 
   redshift $z$ for no change of $m_e$ (solid curve), 
   an increase of $m_e$ by 5$\%$ (dashed curve), a decrease of 
   $m_e$ by 5$\%$ (dotted curve).}
   \label{fig:m_ion_frac}
\end{figure}

\section{Effects on the epoch of recombination and CMB angular power spectrum} \label{sec:CMB}
\begin{figure}
   \centering
   \includegraphics[width=7.5cm]{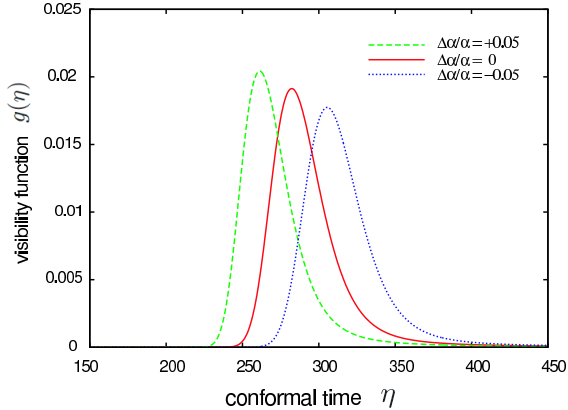}
   \caption{The visibility function as a function of conformal 
   time for no change of $\alpha$ (solid curve), an increase 
   of $\alpha$ by 5$\%$ (dashed curve), a decrease of 
   $\alpha$ by 5$\%$ (dotted curve).}
   \label{fig:a_vis}
\end{figure}

\begin{figure}
   \centering
   \includegraphics[width=7.5cm]{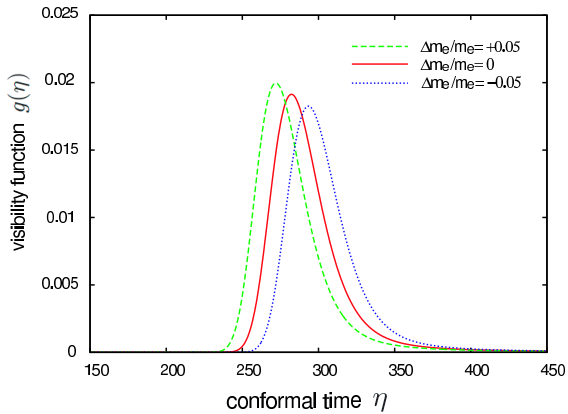}
   \caption{The visibility function as a function of conformal 
   time for no change of $m_e$ (solid curve), an increase of 
   $m_e$ by 5$\%$ (dashed curve), a decrease of $m_e$ by 5$\%$ 
   (dotted curve).}
   \label{fig:m_vis}
\end{figure}

We have investigated the equations which describe the process of recombination and how they depend on the coupling constants in the previous section. We incorporate the dependence on $\alpha$ and $m_e$ into the RECFAST code \cite{Seager:1999bc} and solve the equations for the ionization fraction $x_e$ as a function of redshift with several different values of $\alpha$ and $m_e$. The results are shown in Figs.~\ref{fig:a_ion_frac} and \ref{fig:m_ion_frac}.
We have assumed a flat universe and used cosmological parameters $(\omega_b, \omega_m, h) = (0.024,0.14, 0.72)$, where $\omega_b \equiv \Omega_b h^2$ is the baryon density, $\omega_m \equiv \Omega_m h^2$ is the matter density, $h$ is the Hubble parameter, and $\Omega$ denotes the energy density in unit of the critical density.
The most important feature is the shift of the epoch of recombination
to higher $z$ as $\alpha$ or $m_e$ increases. We can also see this by the rightward shift of the peak of the visibility function shown in Figs.~\ref{fig:a_vis} and \ref{fig:m_vis}. This is easy to
understand because the binding energy $B_n$ scales as $\alpha^2 m_e$ and
photons should have higher energy to ionize hydrogens.

\begin{figure}
  \centering
   \includegraphics[width=7.5cm]{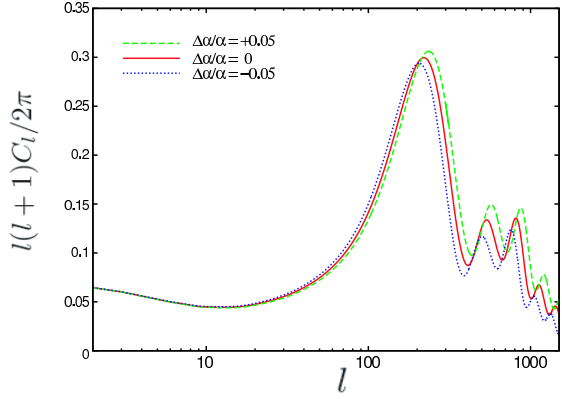}
   \caption{The spectrum of CMB fluctuations for no change 
   of $\alpha$ (solid curve), an increase of $\alpha$ by 5$\%$ 
   (dashed curve), a decrease of $\alpha$ by 5$\%$ (dotted curve).}
   \label{fig:a_cl}
\end{figure}

\begin{figure}
   \centering
   \includegraphics[width=7.5cm]{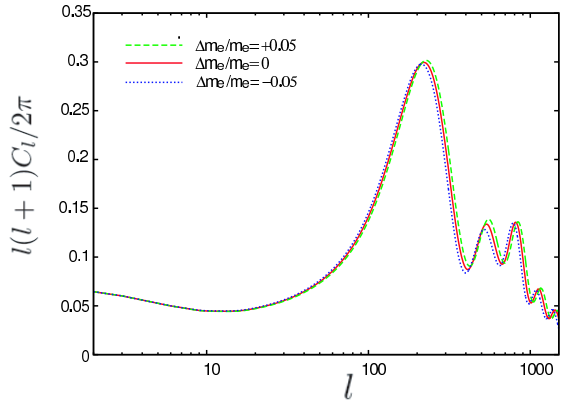}
   \caption{The spectrum of CMB fluctuations for no change 
   of $m_e$ (solid curve), an increase of $m_e$ by 5$\%$ 
   (dashed curve), a decrease of $m_e$ by 5$\%$ (dotted curve).}
   \label{fig:m_cl}
\end{figure}


In Figs.~\ref{fig:a_cl} and~\ref{fig:m_cl} we show
the power spectrum of the CMB temperature anisotropy 
for several different values of $\alpha$ and $m_e$ as calculated by the CMBFAST code \cite{Seljak:1996is} with the modified RECFAST. 
We consider a flat $\Lambda$CDM universe with a power-law adiabatic primordial fluctuation. The adopted cosmological parameter values are $(\omega_b, \omega_m, h,n_s, \tau) = (0.024,0.14, 0.72, 0.99, 0.166)$ where $\tau$ is the reionization optical depth and $n_s$ is the scalar spectral index. We fix the value of the amplitude of primordial power spectrum. We can see two effects of varying $\alpha$ and $m_e$ in Figs.~\ref{fig:a_cl} 
and \ref{fig:m_cl}. Increasing $\alpha$ or $m_e$ shift the peak positions to higher values of $l$ 
and amplify the peak heights.
 
The peak position shift is understood as follows.
Using $l_p$ to denote the position of a peak,
$r_\theta(z)$ for the angular diameter distance and
$r_s(z)$ for the sound horizon, one can write
\begin{equation}
 l_p \sim \frac{r_\theta(z_{ls})}{r_s(z_{ls})},
\end{equation}
where $z_{ls}$ is
the redshift of the last scattering surface. 
Increasing $\alpha$ or $m_e$ increases the redshift of 
the last scattering surface due to the larger binding energy, 
as in Figs.~\ref{fig:a_vis} and~\ref{fig:m_vis}.
The higher $z_{ls}$ in turn corresponds to
a smaller sound horizon and a larger angular diameter distance, 
which lead to a higher value of $l_p$.

The changes in the peak heights are caused by modifications to the early ISW effect and the diffusion damping. The larger $z_{ls}$ leads to the larger early ISW effect making the first peak higher. To consider the effect beyond the first peak, we focus our 
attention on the visibility function. 
The peak of the visibility function moves to a larger 
redshift since the recombination occurs at higher redshift, 
when the expansion rate is faster. Hence, the temperature 
and $x_e$ decreases more rapidly, making the peak width of
the visibility function narrower 
(see Figs.~\ref{fig:a_vis} and \ref{fig:m_vis}).
Since the width of the visibility function corresponds to 
the damping scale, an increase in $\alpha$ or $m_e$ decreases 
the effect of damping. This is the reason why the amplitude at larger $l$ increases with
increasing $\alpha$ and $m_e$. 
Moreover, as seen in Figs.~\ref{fig:a_vis} and \ref{fig:m_vis}, $\alpha$ changes the visibility function width more than $m_e$ does (quantitatively, an increase of $\alpha$ or $m_e$ by 5$\%$
makes the full width at half maximum of the visibility function 
narrower by 10$\%$ or 2$\%$ respectively) because the binding energy which scales as $\alpha^2 m_e$. 
Thus the damping scale is more sensitive to the change of $\alpha$ than $m_e$, as appears in Figs.~\ref{fig:a_cl} and \ref{fig:m_cl}.

Now we discuss the effects of varying $\alpha$ and $m_e$ somewhat more quantitatively using the following four quantities which characterize a shape of
CMB power spectrum \cite{Hu:2000ti}:
the position of the first peak $l_1$, the height of the first peak
relative to the large angular-scale amplitude evaluated at
$l = 10$,
\begin{equation}
        H_1 \equiv \left(\frac{\Delta T_{l_1}}
        {\Delta T_{10}}\right)^2
\end{equation}    
the ratio of the second peak ($l_2$) height to the first
 \begin{equation}
        H_2 \equiv \left(\frac{\Delta T_{l_2}}
        {\Delta T_{l_1}}\right)^2
\end{equation}    
the ratio of the third peak ($l_3$) height to the first
 \begin{equation}
        H_3 \equiv \left(\frac{\Delta T_{l_3}}
        {\Delta T_{l_1}}\right)^2
\end{equation}    
where $(\Delta T_l)^2 \equiv l(l + 1)C_l/2\pi$.
Note that these four quantities do not depend on overall amplitude.
We calculate the response of these four quantities when 
we vary the parameters $\omega_b$, $\omega_m$, $h$,
$\tau$, $n_s$, $\alpha$ and $m_e$.
When we vary one parameter, the other parameters are fixed
and flatness is always assumed 
(especially, increasing $h$ means increasing $\Omega_\Lambda$
because $\Omega_\Lambda = 1 - \omega_m/h^2$).
\begin{eqnarray}   
   \Delta l_1 & = & 
        16\frac{\Delta \omega_b }{\omega_b } 
      - 25\frac{\Delta \omega_m }{\omega_m } 
      - 47\frac{\Delta h}{h} \nonumber \\
    & &+  36\frac{\Delta n_s}{n_s} 
      + 290\frac{\Delta \alpha}{\alpha} 
      + 150\frac{\Delta m_e}{m_e} ,
      \label{eq:first_peak_dependency} \\
   \Delta H_1 & = & 
          3.0\frac{\Delta \omega_b }{\omega_b } 
        - 3.0\frac{\Delta \omega_m }{\omega_m } 
        - 2.2\frac{\Delta h}{h} 
        - 1.7\frac{\Delta \tau}{\tau} \nonumber \\
    & &+ 18\frac{\Delta n_s}{n_s} 
         + 3.9\frac{\Delta \alpha}{\alpha}
         + 1.4\frac{\Delta m_e}{m_e} ,
         \label{eq:height_dependency} \\
     \Delta H_2 &=& -0.30\frac{\Delta \omega_b }{\omega_b } 
        +0.015\frac{\Delta \omega_m }{\omega_m }+ 0.41\frac{\Delta n_s}{n_s} \nonumber  \\
       & &  + 0.91\frac{\Delta \alpha}{\alpha}
         + 0.30\frac{\Delta m_e}{m_e} ,
 \\
     \Delta H_3 &=& -0.19\frac{\Delta \omega_b }{\omega_b } 
        +0.21\frac{\Delta \omega_m }{\omega_m }+ 0.56\frac{\Delta n_s}{n_s} \nonumber  \\
       & &  + 0.57\frac{\Delta \alpha}{\alpha}
         -0.019\frac{\Delta m_e}{m_e} ,
\end{eqnarray}  
and values at the fiducial parameter values are $l_1 = 220$, $H_1 =6.65$, $H_2 =0.442$ and $H_3 =0.449$. Derivatives of $l_1$ and $H_1$ with respect to $\alpha$ and $m_e$ are positive as is expected from the considerations above. Furthermore, $\Delta l_1/\Delta \alpha$ and $\Delta l_1/\Delta m_e$ are much larger than the other derivatives of $l_1$ while $\Delta H_1/\Delta \alpha$ and $\Delta H_1/\Delta m_e$ have relatively similar values to the other derivatives of $H_1$. Since such changes are most effectively mimicked by the change in $h$, it is considered to be the most degenerate parameter with $\alpha$ and $m_e$.
We have seen above that when $\alpha$ or $m_e$ increases, the first peak is enhanced by larger ISW effect and the second or higher peaks are enhanced by smaller diffusion damping. The derivatives of $H_2$ and $H_3$ tell us which effect is important. 
Since $\Delta H_2/\Delta \alpha$ and $\Delta H_3/\Delta \alpha$ are positive and larger than the derivatives with respect to $m_e$, we see that the effect on the diffusion damping is more significant than that on the early ISW for varying $\alpha$. They seem to somewhat cancel each other for varying $m_e$ especially regarding $H_3$. Such behavior is consistent with the consideration at the end of the previous paragraph, that the diffusion damping is more sensitive to the change in $\alpha$ than $m_e$. 


\section{Relation between variations of $\alpha$ and $m_e$} \label{sec:relation}
We expect a unified theory can predict the values of the coupling constants, how they are related to each other and how much they vary in cosmological time scale. 
In string theory, a candidate for unified theory, there is a dilaton field whose expectation value determines the values of coupling constants. However, since it is not fully formulated at present, we have to assume how $\alpha$ and $m_e$ are related to vary and constrain their variations. 
To be concrete, following Ref.~\cite{Campbell:1994bf}, let us start from considering the low energy action 
derived from heterotic string theory in the Einstein frame. The action is written as
\begin{eqnarray}\label{eq:heterotic_action}
        S & = & \int d^4x\sqrt{-g}\left(\frac{1}{2\kappa^2}R 
        - \frac{1}{2}\partial_\mu\Phi\partial^\mu\Phi \right.
        \nonumber \\
        &  - & \frac{1}{2}D_\mu\phi D^\mu\phi - \Omega^{-2}V(\phi) 
        \nonumber \\ 
        &  - & \left.\bar{\psi}\gamma_\mu D^\mu\psi 
        - \Omega^{-1}m_\psi\bar{\psi}\psi 
        - \frac{\alpha'}{16\kappa^2}\Omega^2F_{\mu\nu}F^{\mu\nu}\right), 
\end{eqnarray}
where $\Phi$ is the dilaton field, $\phi$ is an arbitrary 
scalar field, and $\psi$ is an arbitrary fermion. 
$D_\mu$ is the gauge covariant derivative corresponding to 
gauge fields with field strength $F_{\mu\nu}$, 
$\kappa^2 = 8\pi G$ and $\Omega = e^{-\kappa\Phi/\sqrt{2}}$
is the conformal factor which is used to move from string frame.
More concretely, $\phi$ is the Higgs field and $V(\phi)$ is 
its potential.
The overall factor $\Omega$ before the scalar potential
means that the Higgs vacuum expectation value $\langle H \rangle$
is independent of the dilaton so it is taken to 
be constant.
$F_{\mu\nu}$ is the gauge field with gauge group
including ${\rm SU(3)}\times{\rm SU(2)}\times{\rm U(1)}$.
We define its Lagrangian density for the gauge field as 
$-(1/4g^2)F_{\mu\nu}F^{\mu\nu}$
where $g$ is the unified coupling constant.
Compared with equation Eq.~(\ref{eq:heterotic_action}),
\begin{equation}
   \frac{1}{g(M_p)^2} = 
   \frac{\alpha' e^{-\sqrt{2}\kappa\Phi}}{4\kappa^2},
\end{equation}
where $M_p$ is the Planck scale.
We can calculate the gauge coupling constants at low energy
using renormalization group equations.
$\alpha$ almost does not run, and hence the $\alpha$ at low energy
\begin{equation}
   \alpha \simeq \alpha(M_p) 
   = \frac{g(M_p)^2}{4\pi} 
   = \frac{\kappa^2 e^{\sqrt{2}\kappa\Phi}}{\pi\alpha'}.
\end{equation}
As for the other gauge coupling constants, variation of the strong coupling constant may affect CMB since its low energy value determines the QCD scale $\Lambda_{QCD}$ which in turn determines nucleon masses. However, how the variation of $\Lambda_{QCD}$ is related to that of $\alpha$ at low energy can not uniquely be determined from eq.~(\ref{eq:heterotic_action}) and especially depends on the details of unification scheme \cite{Dine:2002ir}. Therefore, for simplicity, we just assume $\Lambda_{QCD}$ does not vary.
The $\psi$'s are the ordinary standard model leptons
and quarks.
As we take $\langle H \rangle = {\rm const}$, 
the Yukawa couplings depend on
the dilaton as $e^{\kappa\Phi/\sqrt{2}}$. 
Therefore the relation between variations of $\alpha$ and $m_e$
is given by
\begin{equation}\label{eq:variations_relation1}
   \frac{m_e +\Delta m_e}{m_e}
    = \left(\frac{\alpha +\Delta \alpha}{\alpha}\right)^{1/2}.
\end{equation}

In this paper, we also consider other possibilities phenomenologically by adopting a power law relation as
\begin{equation}\label{eq:variations_relation}
   \frac{m_e +\Delta m_e}{m_e}
    = \left(\frac{\alpha +\Delta \alpha}{\alpha}\right)^p,
    \end{equation}
and compute constraints for several values of $p$.
In addition to the case with changing only $\alpha$ ($p=0$) and the model described above ($p=1/2$), we consider cases with $p=2$ and 4.

\section{Constraints on varying $\alpha$ and $m_e$} \label{sec:constraint}
We constrain the variation of $\alpha$ in the models described in the previous section using the WMAP first-year data. CMB power spectrum is calculated by CMBFAST \cite{Seljak:1996is} with RECFAST \cite{Seager:1999bc} modified as in Sec.~\ref{sec:rec}. The $\chi^2$ is computed for TT and TE data set by the likelihood code supplied by the WMAP team \cite{Verde:2003ey,Hinshaw:2003ex,Kogut:2003et}. As defined in Sec.~\ref{sec:CMB}, we consider six cosmological parameters $\omega_b$, $\omega_m$, $h$, $\tau$, $n_s$ and overall amplitude $A$ in the $\Lambda$CDM model assuming the flatness of the universe. We report $A$ in terms of $l(l + 1)C_l/2\pi$ at $l=2$ in unit of $\mu$K$^2$. In this paper we do not consider gravity waves, running of the spectral index and isocurvature modes. We calculate $\chi^2$ minimum as a function of $\alpha$ and derive constraints on $\alpha$. The minimization over six other parameters are performed by iterative applications of the Brent method \cite{brent} of the successive parabolic interpolation. More detailed description of this minimization method is found in Ref.~\cite{Ichikawa:2004zi}.
We search for minimum in the region $\tau < 0.3$, which is a prior adopted in Refs.~\cite{Martins:2003pe,Rocha:2003gc}. We derive constraints with or without the constraint on the hubble parameter $h$. When we combine it, we use the Hubble Space Telescope (HST) Hubble Key Project value $h = 0.72 \pm 0.08$ \cite{Freedman:2000cf} whose error is regarded as gaussian 1$\sigma$.

\begin{figure}
   \centering
   \includegraphics[width=7.5cm,clip]{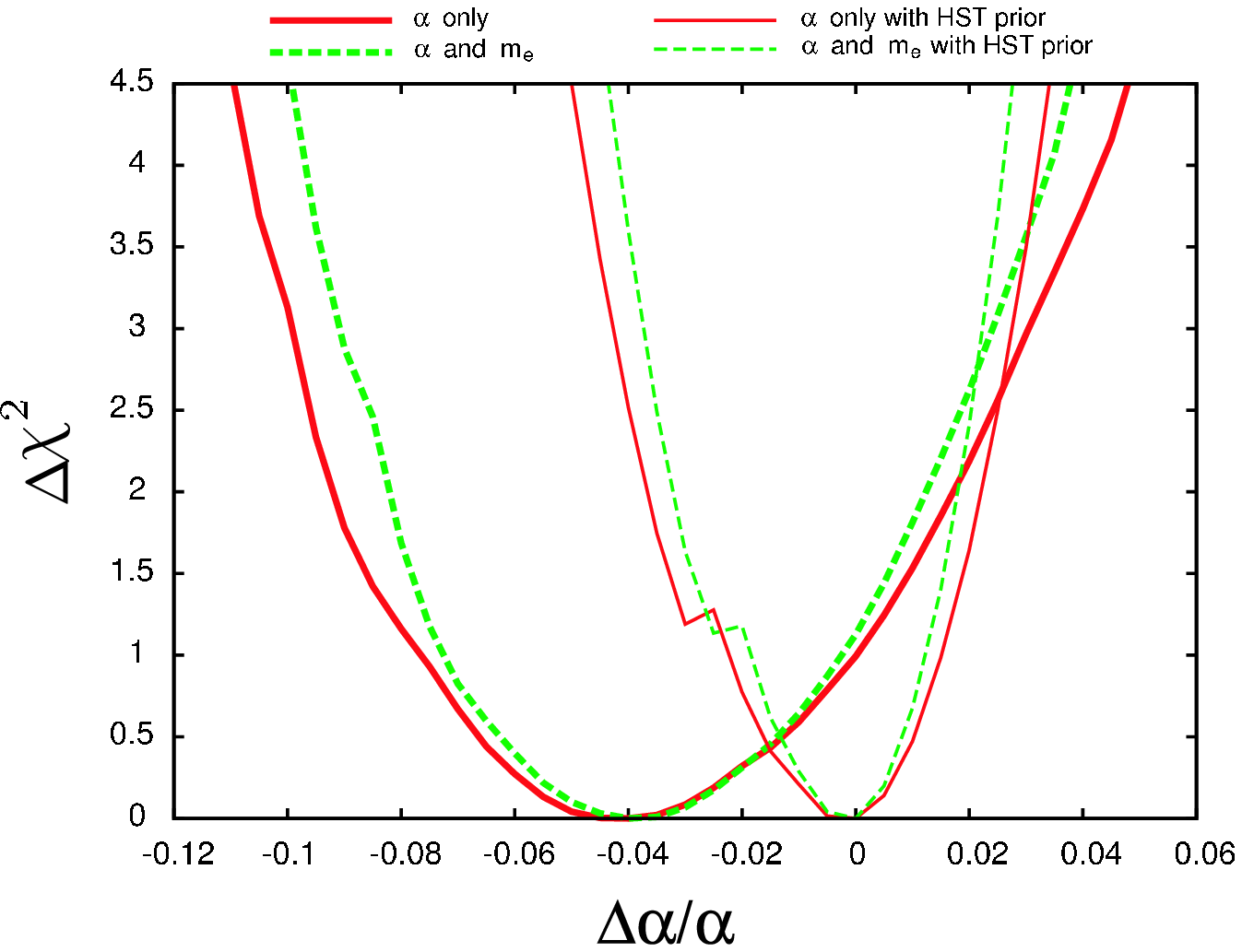}
   \caption{$\chi^2$ function for 
   variation in $\alpha$ only (solid curves) 
   and $\alpha$ and $m_e$ simultaneously with the relation of eq.~(\ref{eq:variations_relation1}) (dashed curves) .
   The HST prior is imposed for the thinner curves.}
   \label{fig:chi2_p}
\end{figure}

Fig.~\ref{fig:chi2_p} shows our results of $\chi^2$ minimization. It compares varying $\alpha$ only and varying $\alpha$ and $m_e$ with the relation of eq.~(\ref{eq:variations_relation1}), respectively with or without the HST prior. Without the HST prior, we find at 95$\%$ C.L. that $-0.107 < \Delta \alpha/\alpha < 0.043$ with changing $\alpha$ only and $-0.097 < \Delta \alpha/\alpha < 0.034$ with the model described in the previous section. Although the best fit $\alpha$ is 4\% less than the present value, we find that $\Delta\alpha =0$ is consistent with the WMAP observation and evidence for varying $\alpha$ is not obtained. The effect of varying $m_e$ simultaneously is found to make the constraint more stringent by 13\%. This rather small effect is reasonable since, as is discussed in Sec.~\ref{sec:CMB}, the effect of $m_e$ on CMB power spectrum is slightly smaller than $\alpha$, and the relation of eq.~(\ref{eq:variations_relation1}) we adopt here does not change $m_e$ much relative to $\alpha$.

\begin{figure}
   \centering
   \includegraphics[width=7.5cm,clip]{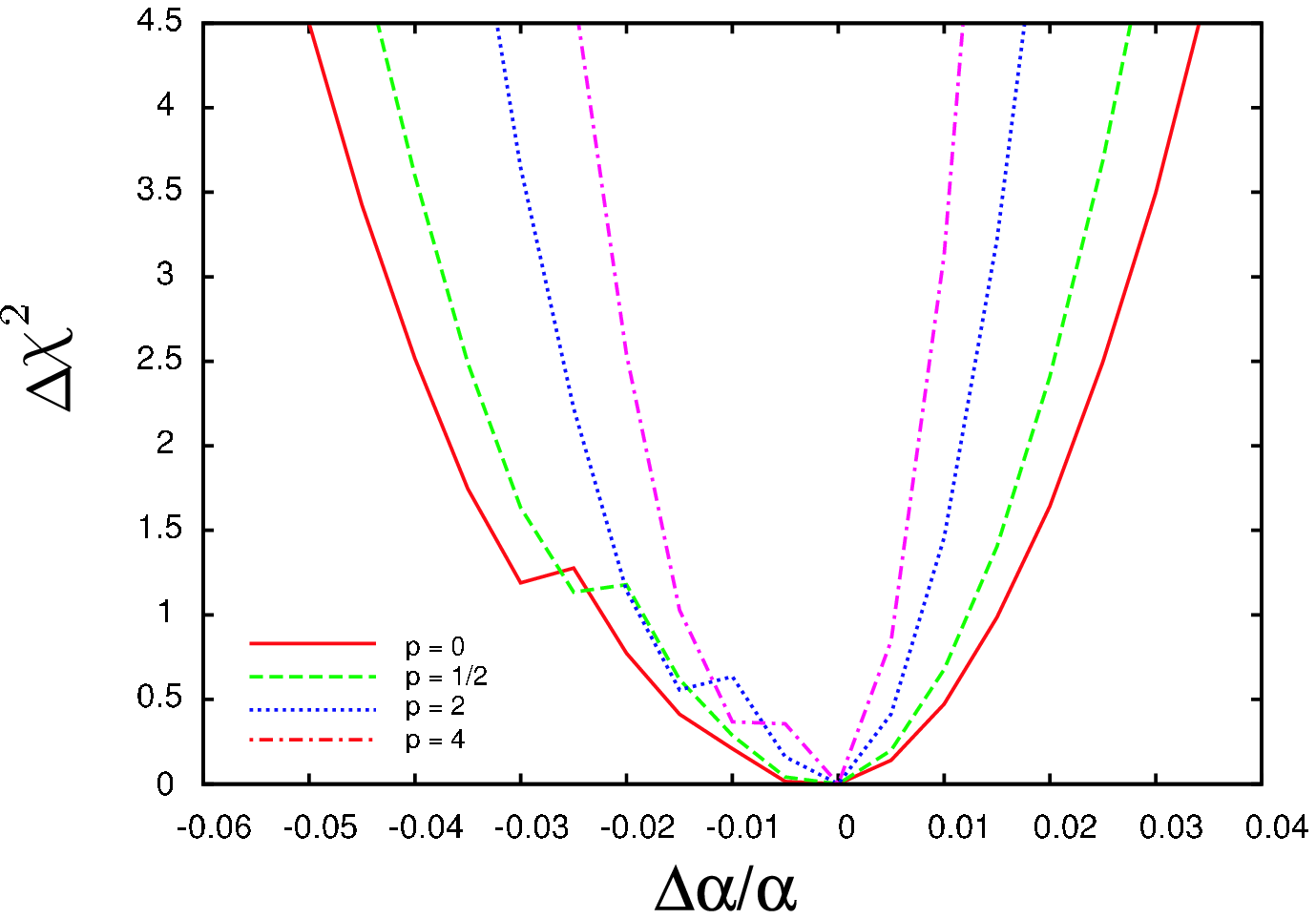}
   \caption{$\chi^2$ function for 
   variation in $\alpha$ and $m_e$ with several values of $p$ defined in Eq.~(\ref{eq:variations_relation}).
   The curves are for, from outside to inside, $p = 0, 1/2, 2$ and 4. The HST prior is imposed.}
   \label{fig:chi2_p_2}
\end{figure}

\begin{table}
   \begin{tabular}{|c|c|} \hline \hline
     power law index & constraint (95$\%$ C.L.) \\ \hline
     $p = 0$ & $-0.048 < \Delta \alpha/\alpha < 0.032$ \\ \hline
     $p = 1/2$ &  $-0.042 < \Delta \alpha/\alpha < 0.026$  \\ \hline
     $p = 2$ &   $-0.031 < \Delta \alpha/\alpha < 0.017$ \\ \hline
     $p = 4$ &   $-0.023 < \Delta \alpha/\alpha < 0.011$ \\ \hline\hline
   \end{tabular}
   \caption{The constraints on $\alpha$ with $m_e$ varies according to various power law relation eq.~(\ref{eq:variations_relation}) of the index $p$. The HST prior is imposed.}
   \label{table:priors}
\end{table}

We find that minimum $\chi^2$ is given at $\Delta \alpha/\alpha = -0.04$ with $(\omega_b, \omega_m, h, n_s, \tau, A) = (0.021, 0.132, 0.523, 0.979, 0.146, 942)$ for the case of  changing $\alpha$ only, and $\Delta \alpha/\alpha = -0.04$ with $(\omega_b, \omega_m, h, n_s, \tau,A) = (0.020, 0.131, 0.485, 0.979, 0.140, 907)$ for the case of changing $\alpha$ and $m_e$ together. Both cases have notably small values of $h$. Since $h$ is considered to be the most degenerate parameter with $\alpha$ or $m_e$ as discussed in the end of Sec.~\ref{sec:CMB}, it is instructive to investigate how constraints tighten when $h$ is limited to higher values such as the HST measurement. From Fig.~\ref{fig:chi2_p}, we obtain, with the HST prior, that 
$-0.048 < \Delta \alpha/\alpha < 0.032$ with changing $\alpha$ only, and $-0.042 < \Delta \alpha/\alpha < 0.026$ with the model described in the previous section. Compared with no HST prior constraints, they are stringent by about factor of 2 for both cases. Moreover, since low values of $h$ which give good fit with $\Delta\alpha/\alpha \approx -0.04$ are ruled out by the HST prior, the center of allowed region has shifted to larger $\Delta \alpha$.

Here, we comment on the constraint previously obtained by Refs.~\cite{Martins:2003pe,Rocha:2003gc} from the WMAP data. As mentioned in Sec.~\ref{sec:intro}, they reported the constraint on $\alpha$ to be $-0.06 < \Delta \alpha/\alpha < 0.01$ (95\% C.L.). They fixed $m_e$ when varying $\alpha$ and values quoted here is the case with no running for the primordial power spectrum. This constraint seems to have been obtained with marginalization on grid with $0 < \Omega_\Lambda < 0.95$ \cite{Martins:2003pe} so it should be compared with our constraint without the HST prior, $-0.107 < \Delta \alpha/\alpha < 0.043$, which is much weaker than theirs. The difference might be traced to the different analysis method but we could not reproduce their results by our method.
 
Finally, we investigate the cases in which $m_e$ varies more than $\alpha$. We consider the models with $p=2$ and 4 in eq.~(\ref{eq:variations_relation}). We calculate constraints with the HST prior and results are summarized in Fig.~\ref{fig:chi2_p_2} and Table~\ref{table:priors} along with $p=0$ and $1/2$ cases. Compared with $p=0$ (only varying $\alpha$) case, the constraints become smaller by 40\% ($p=2$) and 60\% ($p=4$). Although those constraints are much smaller than the case with $p=1/2$, they are still consistent with $\Delta \alpha = 0$.

\section{CONCLUSION} \label{sec:conclusion}

In summary, we have studied a CMB constraint on the time varying 
fine structure constant $\alpha$ taking into account simultaneous change of
electron mass $m_e$ which might be implied in superstring theories. 
We have searched sufficiently wide ranges of the cosmological 
parameters and obtained the WMAP only constraint at 95$\%$ C.L. as
$-0.097 < \Delta \alpha/\alpha < 0.034$.
Combining with the measurement of $h$ by the HST Hubble Key Project, we have obtained more stringent constraint as $-0.042 < \Delta \alpha/\alpha < 0.026$, which improvement is explained by the strong degeneracy between $\alpha$ or $m_e$ and $h$. 
These constraints, obtained by adopting the model with eq.~(\ref{eq:variations_relation1}) are only slightly tighter than those assuming only $\alpha$ variation: $-0.107 < \Delta \alpha/\alpha < 0.043$ and $-0.048 < \Delta \alpha/\alpha < 0.032$, without and with the HST prior respectively.
This is reasonable since the effect of $m_e$ on CMB power spectrum is similar to that of $\alpha$ (sec.~\ref{sec:CMB}) and we adopted to vary $m_e$ milder than $\alpha$ ( eq.~(\ref{eq:variations_relation1})). 

We have also considered other possibilities for the relation between $\Delta \alpha$ and $\Delta m_e$ as in the form of eq.~(\ref{eq:variations_relation}) with $p=2$ and 4. In these cases, the constraints become tighter by roughly a factor of two. This may not look as drastic as the BBN bounds but CMB bounds are promising and have advantage that there will be future experiments with higher sensitivities to $\alpha$ as investigated in Refs.~\cite{Martins:2003pe,Rocha:2003gc}. In this paper, we do not find evidence of varying $\alpha$ in the CMB data of WMAP. We will wait and see whether future experiments give us more stringent bound on $\alpha$ or evidence for varying $\alpha$.


\end{document}